\documentclass[runningheads]{llncs}
\usepackage{graphicx}
\usepackage{amsmath}
\usepackage{amsfonts}
\usepackage{color}
\usepackage{hyperref}
\newcommand{\bbC}[0]{\mathbb{C}}

\begin{document}
\title{Accelerated Cardiac Parametric Mapping using Deep Learning-Refined Subspace Models}
\titlerunning{Deep Learning-Refined Subspace Models}
\date{October 12, 2023}
\author{Calder D. Sheagren\inst{1,2,*}\orcidID{0000-0001-7439-2906} \and \\
Brenden T. Kadota\inst{1,2}\orcidID{0009-0001-1510-7746} \and \\
Jaykumar H. Patel\inst{1,2}\orcidID{0000-0003-1954-4879}
\and \\ 
Mark Chiew\inst{1,2}\orcidID{0000-0001-6272-8783} \and \\
Graham A. Wright\inst{1,2}}
\authorrunning{C. Sheagren et al.}
\institute{$^1$Department of Medical Biophysics, University of Toronto, Toronto, Ontario, Canada 
\\ $^2$Physical Science Platform, Sunnybrook Research Institute, Ontario, Canada \\
*Corresponding email: \email{calder.sheagren@mail.utoronto.ca}}
\maketitle              % typeset the header of the contribution
\begin{abstract}
Cardiac parametric mapping is useful for evaluating cardiac fibrosis and edema. Parametric mapping relies on single-shot heartbeat-by-heartbeat imaging, which is susceptible to intra-shot motion during the imaging window. However, reducing the imaging window requires undersampled reconstruction techniques to preserve image fidelity and spatial resolution. The proposed approach is based on a low-rank tensor model of the multi-dimensional data, which jointly estimates spatial basis images and temporal basis time-courses from an auxiliary parallel imaging reconstruction. The tensor-estimated spatial basis is then further refined using a deep neural network, trained in a fully supervised fashion, improving the fidelity of the spatial basis using learned representations of cardiac basis functions. This two-stage spatial basis estimation will be compared against Fourier-based reconstructions and parallel imaging alone to demonstrate the sharpening and denoising properties of the deep learning-based subspace analysis. 

\end{abstract}
\section{Introduction}

$T_1$-weighted ($T_1$w) and $T_2$-weighted ($T_2$w) imaging are staple techniques to determine regions of cardiac muscle, fat, fibrosis, edema, and amyloid, among other possible pathologies \cite{henningsson_black-blood_2022}. However, these are qualitative imaging techniques that rely on hyperintense or hypointense signal intensities to make a diagnosis, the distribution of which can change as a function of acquisition parameters and sequence type. Moreover, diagnosing diffuse disease on contrast-weighted imaging is difficult due to the global impact of the pathology on myocardial signal intensity. To address these issues, parametric mapping has been introduced to reproducibly quantify the underlying tissue parameters $T_1$ and $T_2$ at every voxel, providing absolute normative and abnormal values for a given field strength.  

Cardiac $T_1$ and $T_2$ mapping have been used clinically to diagnose diseases such as acute myocardial infarction, cardiac amyloidosis, Fabry's disease, and iron overload, among other diseases \cite{messroghli_myocardial_2007}. Additionally, $T_1$ and $T_2$ mapping play an integral role in diagnosing acute myocarditis as per the Lake Louise criteria \cite{luetkens_comparison_2019}. One challenge with parametric mapping is the long temporal window needed to acquire single-shot contrast-weighted images in the cardiac cycle. This causes a trade-off between intra-shot cardiac motion, spatial resolution, and signal-to-noise ratio (SNR) \cite{sheagren_motion-compensated_2023}. 
To shorten the temporal footprint without sacrificing SNR, parallel imaging techniques such as SENSE and GRAPPA that exploit redundancies between the multi-coil acquisitions have been introduced to accelerate acquisitions by a factor of 2-3 \cite{pruessmann_sense_1999,griswold_generalized_2002}. Parallel imaging is widely used in the clinic, but has a practical limit to the acceleration factors supported. To further reduce the temporal window and enable imaging at high heart rates, imaging at higher parallel imaging acceleration factors is needed, requiring more sophisticated undersampled data reconstruction methods to preserve SNR and image quality. 

Parametric mapping can be interpreted as a \emph{spatiotemporal} imaging method, where the same spatial volume is imaged at multiple timepoints with different contrasts. In spatiotemporal imaging, both the space and time dimensions are often highly compressible \cite{feng_xd-grasp_2016,ong_extreme_2020}. One way to leverage this compressibility is to explicitly factor the spatiotemporal image $X\in\bbC^{N_x \times N_y \times N_t}$ as a single outer product $X = C\otimes T$. Here, $C \in \bbC^{N_x\times N_y}$ is a \emph{spatial basis}, and $T \in \bbC^{N_t}$ is a \emph{temporal basis}. This model is very simplistic and may not capture complex interactions between physiology and motion or contrast changes, so the notion of \emph{partially separable functions} has been introduced to model $X$ as a linear combination \begin{equation}
  X=\sum_{l=1}^La_lC_l\otimes T_l.
\end{equation}
Now, each basis component $C_l\in\bbC^{N_x\times N_y}$, $T_l\in\bbC^{N_t}$, for $l = 0, \ldots, L\in\mathbb{N}$, where $L$ is the chosen rank of the tensor. Decomposing spatial information from temporal dynamics allows for sophisticated reconstruction methods with dimension-specific regularizers, such as spatial sparsity preservation and temporal low-rank preservation \cite{zhang_high-resolution_2016}. 

To generalize dimension-specific regularizers, the notion of \emph{deep subspace learning} was introduced to apply deep learning to sub-components of the reconstructed tensor to preserve temporal dynamics while enhancing spatial information \cite{chen_deep_2019}. In this paper, we propose a method using Deep learning-Refined sUbspace ModelS (DRUMs). 

\section{Methods}

\begin{figure}[t]
    \centering
    \includegraphics[width=\textwidth]{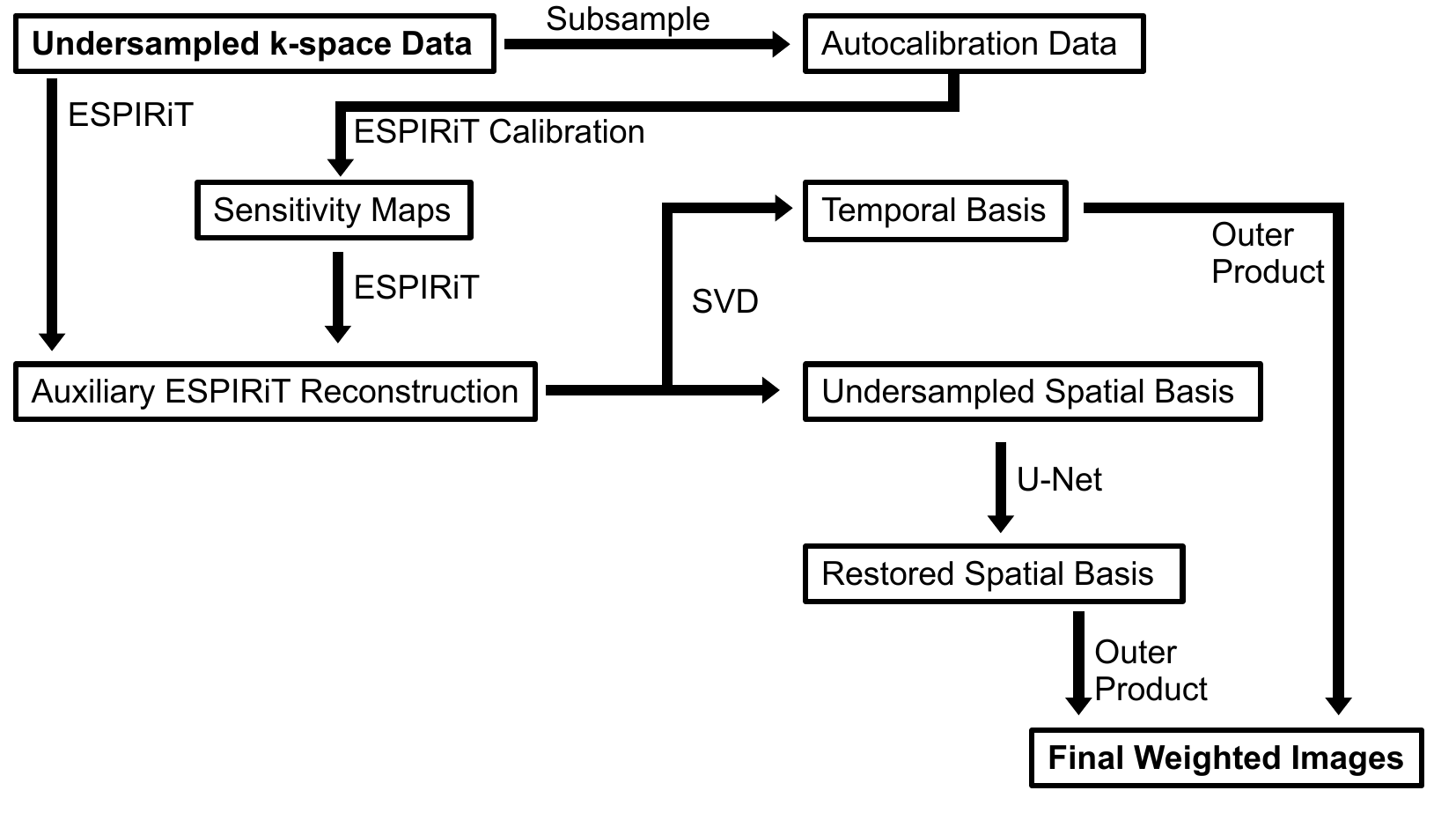}
    \caption{Visual overview of proposed method. The undersampled k-space data is subsampled into the autocalibration data, which is used for sensitivity map calculation. Undersampled k-space data is used to calculate coil sensitivity maps, which jointly are inputs to an auxiliary ESPIRiT reconstruction. The SVD of the ESPIRiT reconstruction is used to generate spatial and temporal basis sets. The spatial basis is fed through a U-Net designed to remove residual noise and incorporate missing high-order image features. The final basis is combined with the temporal basis to obtain the final weighted images. ESPIRiT: \textit{eigenvalue-based iterative self-consistent parallel imaging reconstruction}; SVD: \textit{singular value decomposition}.}
    \label{fig::flow-chart}
\end{figure}

\subsection{Dataset}

The CMRxRecon challenge dataset was used to develop this method \cite{wang_recommendation_2021}. Briefly, 300 volunteers were scanned on a 3 Tesla MRI (MAGNETOM Vida, Siemens Healthineers, Germany). Steady-state free precession $T_1$ maps and gradient-echo $T_2$ maps were acquired with parallel imaging undersampling factors of 2 (hereafter referred to as ``fully sampled"), 4, 8, and 10 to reduce the temporal acquisition window during the cardiac cycle. Sequence parameters for $T_1$ mapping include: FOV = $360\times 307$ mm$^2$, spatial resolution = $1.4 \times 1.4$ mm$^2$, slice thickness = 5.0 mm, TR/TE = 2.67 ms/1.13 ms, Partial Fourier = 7/8, 24 autocalibration lines, TI=\{100, 180, 260, 900, 1000, 1050, 1700, 1800, 2500\}ms. Sequence parameters for $T_2$ mapping include: FOV = $360 \times 288$ mm$^2$, spatial resolution = $1.9 \times 1.9$ mm$^2$, slice thickness = 5.0 mm, TR/TE = 3.06 ms/1.29 ms, T2 preparation times = $\{0, 35, 55\}$ ms, Partial Fourier = 6/8, 24 autocalibration lines. 

\subsection{DRUMs Algorithm}
\begin{figure}[t]
    \centering
    \includegraphics[width=\textwidth]{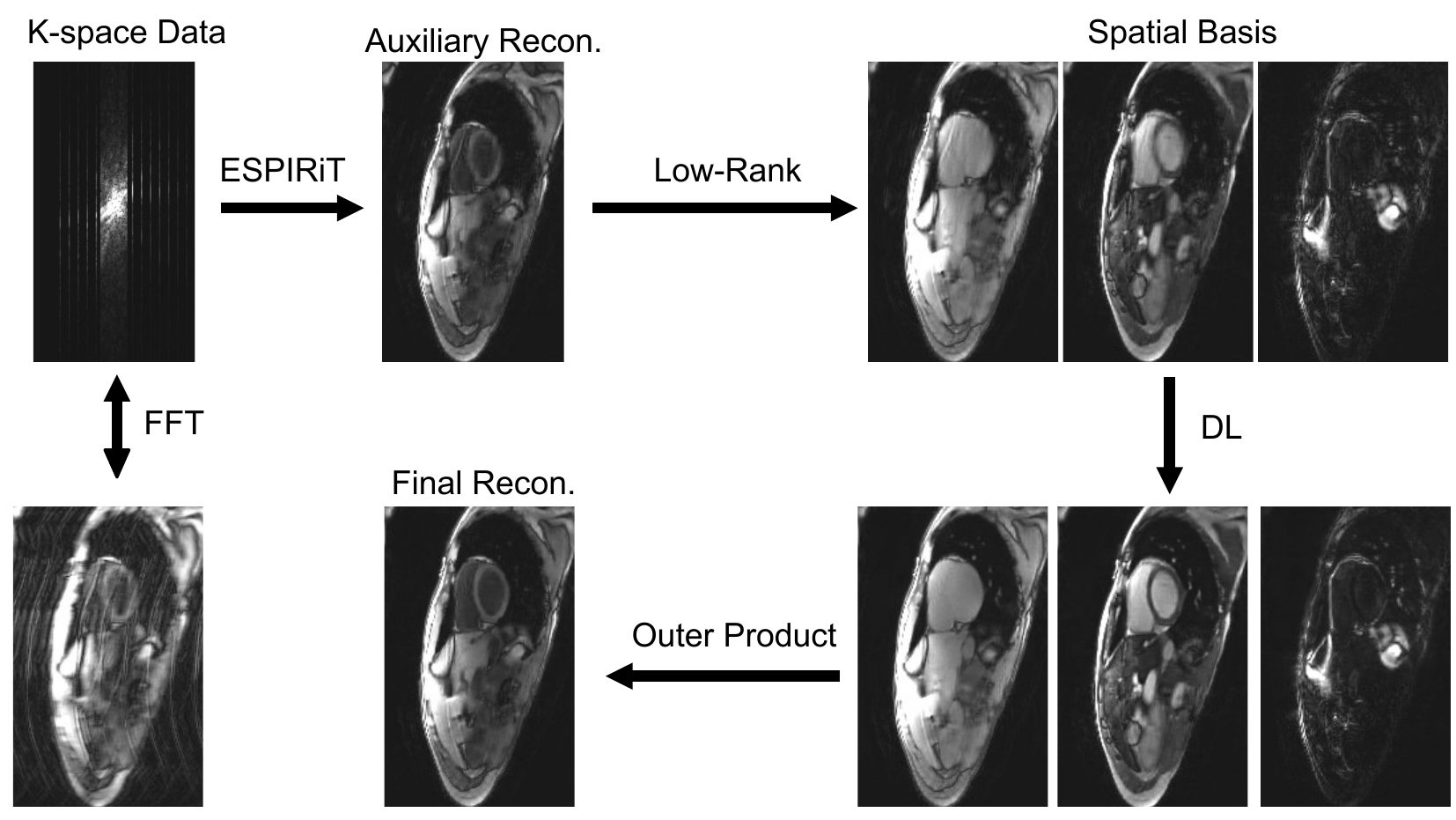}
    \caption{Sample images at multiple stages of the reconstruction pipeline. Raw k-space data is reconstructed using the ESPIRiT method, and is subsequently decomposed into a spatial and temporal basis. A deep learning network is applied to the spatial basis, which is combined with the temporal basis vectors and singular values to obtain a final reconstruction. The FFT reconstruction is provided in the lower left-hand corner for visual comparison with the proposed method. ESPIRiT: \textit{eigenvalue-based iterative self-consistent parallel imaging reconstruction}; FFT: \textit{fast Fourier transform}.}
    \label{fig::sample-images}
\end{figure}

In this section, we will discuss the methods used for the DRUMs reconstruction combining parallel imaging, subspace models, and deep learning-refined spatial bases. For a visual overview of the method, see Fig. \ref{fig::flow-chart}. For sample images at multiple stages in the method, see Fig. \ref{fig::sample-images}. 
Training and inference code for this model is available at \href{https://github.com/WrightGroupSRI/CMRxRecon}{https://github.com/WrightGroupSRI/CMRxRecon}. 

% auxiliary espirit reconstruction on weighted images
One set of coil sensitivity maps $\{S_q\}_{q=1}^{10}$ per slice were calculated using the \verb|bart| ESPIRiT calibration (\verb|ecalib|) utility \cite{uecker_espiriteigenvalue_2014}. These sensitivity maps were used for a slice-by-slice, contrast-independent $L^1$-ESPIRiT reconstruction in the \verb|bart| framework that solves the following inverse problem:
\begin{equation}
    X_E = \arg\min_X{1\over 2}\sum_{q=1}^{10}\left\|PFS_qX - y\right\|_2^2 + \lambda \|\Psi X\|_1.
\end{equation}
Here, $X_E$ is the reconstructed image, $P$ is the undersampling projection mask operator, $F$ is the Fourier transform, $y$ is the undersampled multi-coil k-space data, $\lambda=0.01$ is the regularization parameter determined via visual analysis, and $\Psi$ is the spatial wavelet transform. Sparse thresholding in the wavelet domain was selected as a regularizer due to its performance in prior compressed-sensing applications \cite{lustig_sparse_2007}. 100 iterations were used with an eigenvalue-guided step size selection. 

% spatial basis fitting
$X_E$ was then decomposed into a spatial basis $C$ and a temporal basis $T$ in a slice-by-slice manner using the singular value decomposition. The rank-9 basis for $T_1$ mapping images was truncated to a rank-3 basis, allowing for a constant rank $L=3$ for both $T_1$ and $T_2$ mapping. $T_2$ mapping used the full rank of 3 to improve parametric accuracy at the cost of some spatial regularization.

To recover image sharpness lost in the compressed sensing reconstruction and reduce residual statistical noise gained in the ESPIRiT reconstruction, the spatial basis $C$ was input into a deep learning model $U_\theta$. The network was trained to optimize the model weights $\theta$ in a fully supervised manner using spatial bases generated from undersampled and fully sampled data. For more details, see Section \ref{sec::model-training}. After the model was applied, final weighted images were calculated as 

\begin{equation}
    X = \sum_{l=1}^La_l\left(U_\theta(C_l)\right)\otimes T_l.
\end{equation}

Parameter maps were fitted during the submission validation process using CMRxRecon-specific code. Briefly, $T_1$ maps were fitted using a two-step process:

\begin{equation}
    X({\rm TI}) = A - B\exp(-{\rm TI} / T_1^*), \quad T_1 = \left({B\over A} - 1\right)T_1^*.
\end{equation}
$T_2$ maps were fitted using a one-step process:
\begin{equation}
    X(T_{\rm prep}) = A\exp(-T_{\rm prep} / T_2),
\end{equation}
where $T_{\rm prep}$ denotes the $T_2$ preparation time. 

\subsection{Model Training}\label{sec::model-training} % Please provide some feedbcck. Not sure if I should describe what a U-Net is or just state we used a U-net
A U-Net model architecture was used here due to its success in multiple reconstruction and denoising applications~\cite{lehtinen2018noise2noise,DBLP:journals/corr/RonnebergerFB15}. The purpose of the U-Net in this application is to restore high-order spatial information, reduce statistical noise, and reduce aliasing artifacts in the undersampled spatial basis vectors. Our U-Net consists of 4 downsampling and upsampling layers and initializes with 64 convolutional filters. Each downsampling layer level downsamples the spatial features by a factor of 2 and doubles the convolutional filters. The convolutional kernels are sized at $3 \times 3$ and were followed by Batch normalization and a rectified linear unit activation function. Spatial basis vectors were split into the real and imaginary components of the complex signal, dephased to a consistent phase, and cropped to a constant size of (128, 128). Basis vectors were concatenated into an image size of $(128, 128 \times 2L)$ to pass all three vectors to the model simultaneously.

Our U-Net learns to minimize the residual between the undersampled spatial basis and fully sampled spatial basis coefficients by varying model weights $\theta$:
\begin{equation}
    \hat \theta = \arg\min_{\theta} \mathcal{L}(U_{\theta}(C) - C, \hat{C}).
\end{equation}
Here, $\hat C$ is the fully sampled spatial basis coefficients. We pass spatial basis coefficients with vectors concatenated along the channel dimension through the network. The dataset was normalized using $z$-standardization along the channel dimension for undersampled and fully sampled spatial basis coefficients. Here we use a combination of $L^1$ and Structural Similarity Index (SSIM) losses to help preserve edges in our spatial bases. $L^1$ loss was chosen to enforce voxelwise consistency with the ground-truth data, and the SSIM loss was chosen due to its versatility in image information and contrasts. The total loss function is  
\begin{equation}
    \mathcal{L} = 1 - {\rm SSIM}\left(U_\theta(C), \hat{C}\right) + \|U_\theta(C) - \hat{C}\|_1.
\end{equation}
Here, a window size of 5 was used in the SSIM loss. 

The U-Net was implemented using the PyTorch deep learning library. The given training data was subdivided. We split the training data into internal training, internal validation, and internal test set sizes of 100, 10 and 10 patients (3252, 318, and 318 slices). The Adam optimizer was used with default parameters and a learning rate of $10^{-3}$ with a batch size of 16 slices. The learning rate was chosen using a learning rate range test as shown in \cite{smith_cyclical_2017}. 
Dropout of 0.50 was used to regularize the model to prevent overfitting. A single model for use on all image contrasts and acceleration factors was trained for 500 epochs on an NVIDIA P100 Pascal GPU with 12GB of memory, which took approximately 10 hours. For an overview of the model information, please see Table \ref{tab::model-information}.

\begin{table}[!t]
    \centering
    \begin{tabular}{|l|l|l|l|}
    \hline
       \textbf{Task of Participation}   & Mapping & \textbf{Pre-Training}      & None\\
       \hline
       \textbf{University/Organization} & University of Toronto        & \textbf{Data Augmentation} & None \\
        & Sunnybrook Research Institute        &  &  \\
       \hline
       \textbf{Coil Configuration} & Multi-channel & \textbf{Data Standardization} & Z-score \\
       \hline
       \textbf{Training Hardware} & NVIDIA P100 Pascal 12GB& \textbf{Parameter Number} & 31,036,800\\
       & Intel E5-2683v4 4$\times$2.1GHz &  & \\
       \hline
       \textbf{Validation Hardware} & NVIDIA RTX 2060 Super 8GB &  & \\
       & Intel i7-6700k 6$\times$4.00GHz&& \\
       \hline
       \textbf{Training Time} & 10 hours & \textbf{Loss Function} & $L^1 + {\rm SSIM}$ \\
       \hline
       \textbf{Inference Time} & 3 hours & \textbf{Physical Model} & None (Low-rank) \\
       \hline
       \textbf{Training Set Performance} & See Section \ref{sec::result} & \textbf{Use of Unrolling} & None \\
       \hline
       \textbf{Validation Set Performance} & See Table \ref{tab::validation-set}& \textbf{k-space Fidelity} & Compressed sensing\\
       \hline
       \textbf{Docker Submitted} & Yes & \textbf{Model Backbone} & U-Net  \\
       \hline
       \textbf{Segmentation Labels} & None & \textbf{Operations} & Dephased complex-valued \\
       \hline
    \end{tabular}
    \caption{Description of model information for mapping submission.}
    \label{tab::model-information}
\end{table}

\subsection{Experiments}
Normalized root mean squared error (NRMSE) was compared between fully-sampled images and reconstructed images for patients in the internal testing subset of the overall training dataset with provided fully-sampled reference images. This allows for comparison of what reconstruction steps impacted NRMSE the most and what performance differences are present between acceleration factors and image contrasts. 

Peak signal to noise ratio (PSNR), normalized mean squared error (NMSE), and structural similarity index metric (SSIM) were measured in the validation set between reconstructed images and unseen fully-sampled images. Comparison was performed on the parameter level, with $T_1$ and $T_2$ maps serving as inputs for reconstructed and fully-sampled images. Parameter values were measured over a manually-segmented region of interest containing the left ventricular myocardium, left ventricular blood pool, and right ventricular blood pool.  Results for the DRUMs algorithm were compared against conventional parallel imaging reconstruction as submitted by CMRxRecon on the validation set leaderboard. 

% Results for testing set

\section{Results}\label{sec::result}
The proposed DRUMs method was successfully implemented in Python with a trained neural network. On a CPU, each 3D k-space dataset reconstruction took 100 seconds, which is broken down into the following: Sensitivity map calculation - 23.35s; ESPIRiT reconstruction -  75.96s; SVD and basis fitting - 0.34s; and U-Net inference -  0.33s. On a GPU, the entire training dataset was able to be reconstructed within 5 hours. 

\begin{figure}[t]
    \centering
    \includegraphics[width=\textwidth]{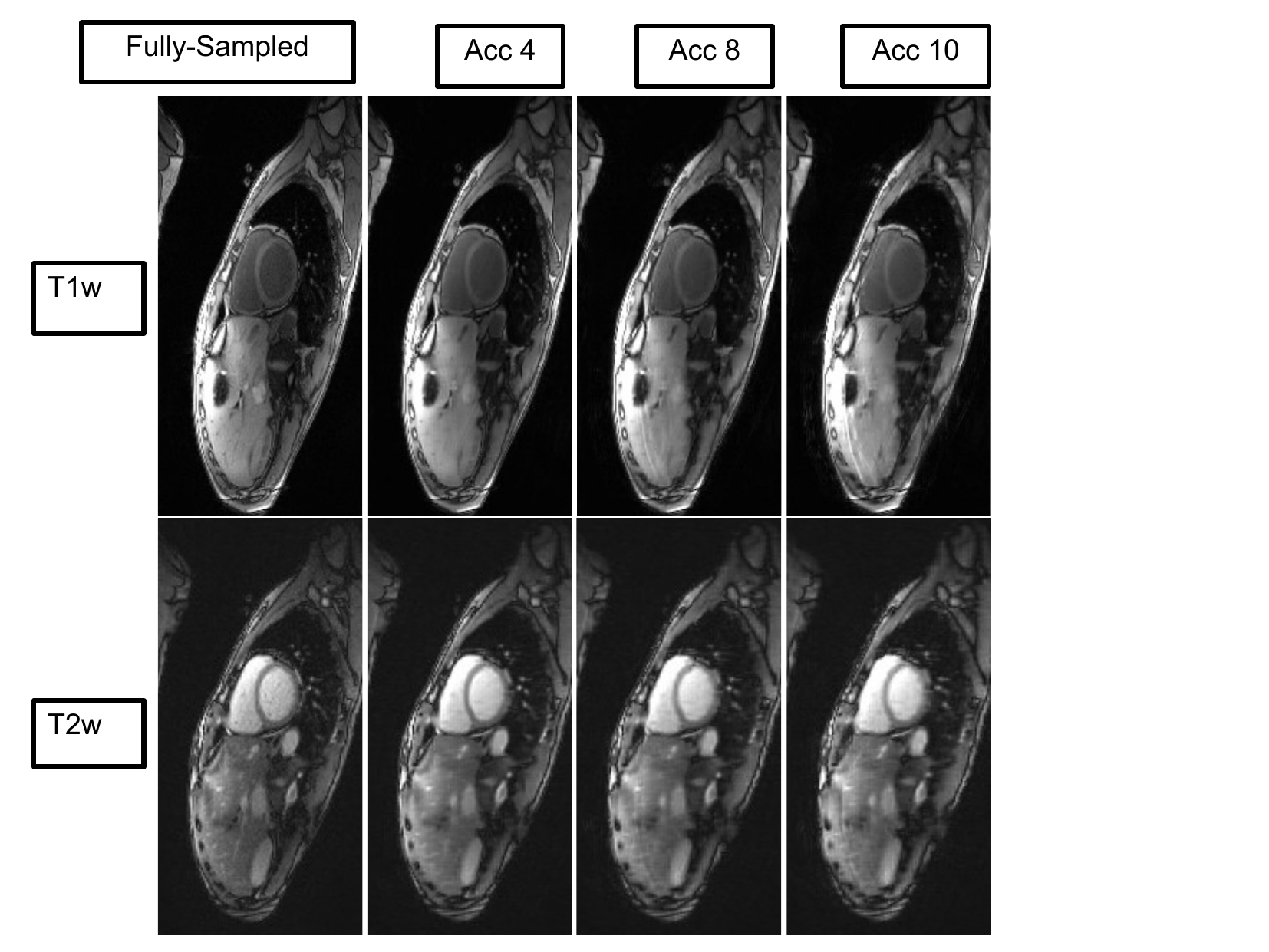}
    \caption{Sample $T_1$w and $T_2$w images reconstructed using the proposed DRUMs method across acceleration factors. Top row: $T_1$w images. Bottom row: $T_2$w images. Left column: fully sampled images reconstructed using the ESPIRiT method to preserve signal intensity scaling. Right columns: images reconstructed at acceleration factors of 4, 8, and 10. Images at higher acceleration factors are visually similar to fully-sampled images, with residual blurring and aliasing artifacts more present at higher acceleration factors, particularly in the $T_2$w images. ESPIRiT: \textit{eigenvalue-based iterative self-consistent parallel imaging reconstruction}}
    \label{fig::accel-comparison}
\end{figure}

For sample images at various acceleration factors, refer to Fig. \ref{fig::accel-comparison}. NRMSE across all contrasts for the zero-filled Fast Fourier Transform (FFT) reconstruction was $0.081\pm 0.02$. NRMSE across all contrasts for the ESPIRiT reconstruction was $0.033 \pm 0.02$. NRMSE across all contrasts for the low-rank approximation of the ESPIRiT reconstruction was $0.035 \pm 0.02$, and NRMSE across all contrasts for the full DRUMs reconstruction was $0.030 \pm .01$. We speculate that the low-rank approximation increased the NRMSE due to constraining the data in a subspace. NRMSE for $T_1$w images was $0.030 \pm 0.02$, and NRMSE for $T_2$w images was $0.028 \pm 0.01$, showing no large difference between the two tissue contrasts. NRMSE for acceleration factors 4, 8, and 10 was $0.023 \pm 0.01$, $0.031 \pm 0.02$, and $0.035 \pm 0.02$, respectively. This is expected, as errors increase with fewer given k-space lines. 

\begin{table}[t]
    \centering
    \begin{tabular}{|l|l|l|l|}
    \hline
    Metric & $R=4$ & $R=8$ & $R=10$  \\
    \hline
    $T_1$ PSNR [dB] & 23.00/\textbf{31.28} & 22.38/\textbf{29.10} & 22.23/\textbf{27.74}  \\
    $T_2$ PSNR [dB] & 23.89/\textbf{29.45} & 23.40/\textbf{27.70} & 23.50/\textbf{27.03} \\
    \hline
    $T_1$ SSIM [1] & 0.66/\textbf{0.85} & 0.63/\textbf{0.81} & 0.63/\textbf{0.79} \\
    $T_2$ SSIM [1] & 0.77/\textbf{0.87} & 0.75/\textbf{0.84} & 0.77/\textbf{0.83} \\
    \hline
    $T_1$ NMSE [1] & 0.21/\textbf{0.03} & 0.26/\textbf{0.05} & 0.27/\textbf{0.06} \\
    $T_2$ NMSE [1] & 0.08/\textbf{0.02} & 0.09/\textbf{0.03} & 0.09/\textbf{0.03} \\
    \hline
    \end{tabular}
    \caption{Validation set performance on the segmented myocardial ROI. In each cell, the leftmost value denotes the CMRxRecon parallel imaging submission, and the rightmost value denotes our proposed DRUMs method. Numbers in bold refer to best-performing methods. $R$: acceleration factor.}
    \label{tab::validation-set}
\end{table}

\begin{figure}[t]
    \centering
    \includegraphics[width=\textwidth]{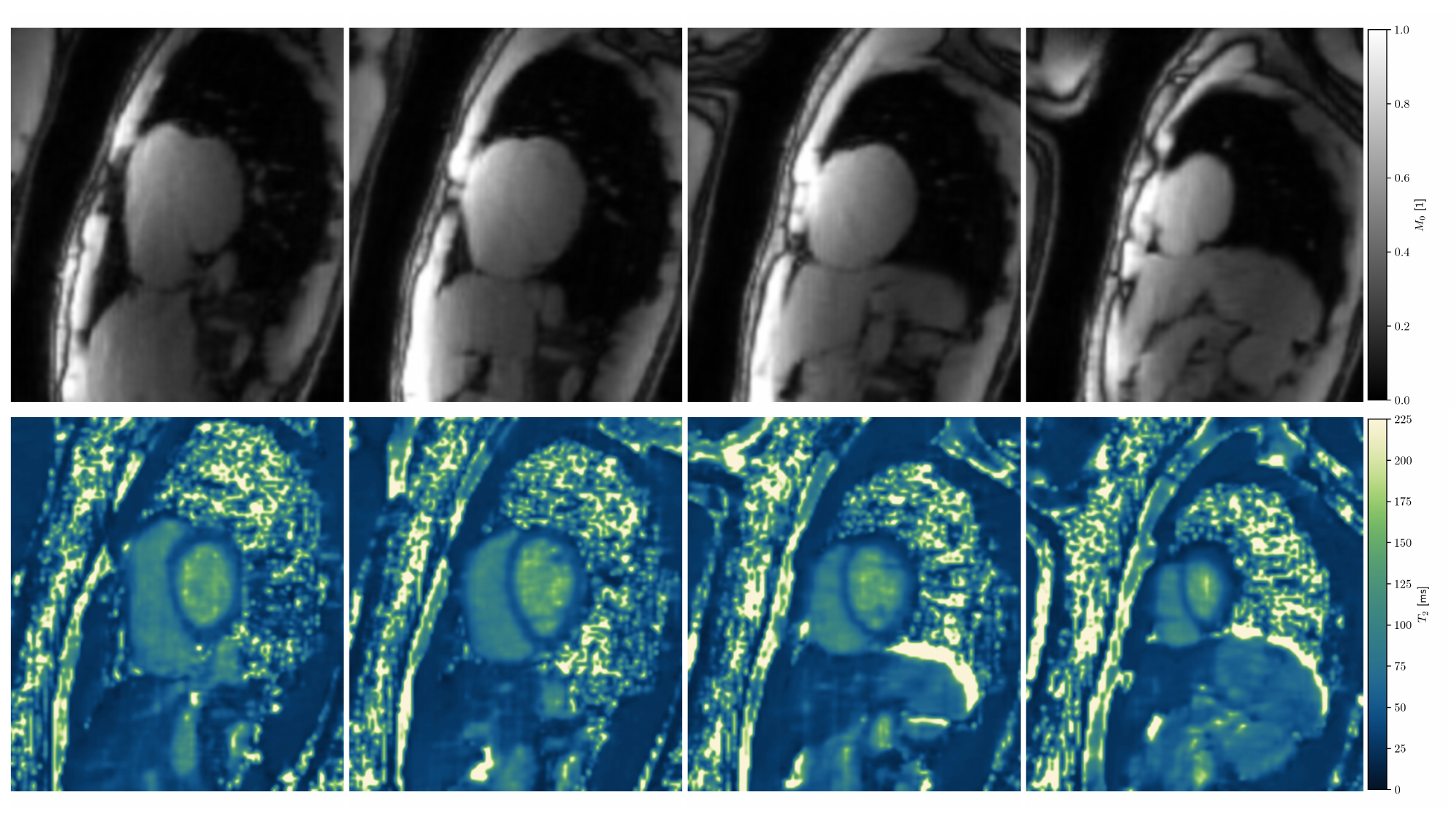}
    \caption{Sample quantitative results for DRUMs reconstructed images. Results from a single subject from the validation set with no provided ground-truth images are plotted here, with a bicubic interpolation for voxel smoothing. Reconstructed images at an acceleration factor of 10 were fitted to the following equation: $S(t) = M_0\exp(-T_{\rm prep}/T_2)$, $M_0\in[0, 1]$, $T_2\in [0, 250]$ms. Top row: proton-density $M_0$ maps. Bottom row: $T_2$ maps. Parameter values are reasonable for 3 Tesla acquisition, and good delineation between myocardium and LV blood pool is observed. The RV free wall is visible in some slices, and residual blurring and aliasing artifacts are present around the heart.}
    \label{fig::t2-example}
\end{figure}

In the unseen validation set, metrics were evaluated using the fitted $T_1$ and $T_2$ maps in a manually-segmented myocardial ROI. Table \ref{tab::validation-set} contains a comparison of the proposed DRUMs method and the benchmark CMRxRecon parallel imaging reconstruction submission. DRUMs reconstruction outperformed the parallel imaging reconstruction in every metric at every acceleration factor. Our method generally performed better quantitatively in $T_2$ reconstruction than $T_1$ reconstruction, with the exception of PSNR, which is surprising given the lack of temporal regularization in our $T_2$ reconstruction. The parallel imaging reconstruction was also superior in $T_2$ mapping vs $T_1$ mapping with the same method for all metrics including PSNR. For sample $M_0$ and $T_2$ maps at acceleration 10 from a patient in the validation set with unseen fully sampled data, see Fig. \ref{fig::t2-example}. 

% Unseen submission data

\section{Discussion and Conclusion}

In this paper, we have proposed a deep learning-refined subspace model reconstruction framework that is compatible with cardiac $T_1$ and $T_2$ mapping. The subspace formalism and constant rank allows for simple cross-application between $T_1$ and $T_2$ mapping. Temporal consistency is enforced using data-driven constraints rather than fitting to predefined modelling functions. Applying the deep learning model to the spatial basis vectors allows for improvements in spatial fidelity while reducing the capability for hallucinating due to the predetermined temporal constraints. 

This proposed method has several limitations. First, it was only trained on one $T_1$ mapping sequence and one $T_2$ mapping sequence from a single vendor, so while we hope this method can generalize well to unknown sequences or contrasts due to its low-rank nature, this has not been rigorously evaluated in this submission. Second, the $T_2$ mapping uses the full rank of 3, which improves parametric accuracy at the cost of temporal compressibility. This can be observed in Fig. \ref{fig::accel-comparison}, where $T_1$w images have improved image quality at higher acceleration values. We will investigate the effects of using rank-2 approximations of $T_2$w multi-contrast images on overall image quality and parametric robustness. Finally, the choice of parallel imaging undersampling pattern was not optimized for temporal incoherence of artifacts. Temporally-incoherent artifacts improve the performance of sparsity-based reconstruction methods like compressed sensing, and can be accomplished using time-varying pseudorandom undersampling or variable-density sampling patterns \cite{ahmad_variable_2015}. Generally, choosing different higher-order k-space lines at different contrasts allows for more robust temporal regularization that can admit higher acceleration factors. 

In the future, we hope this method can be applied inline on scanners to produce high-quality reconstructions in accelerated sequences that facilitate greater use of advanced reconstruction methods in standard clinical practice. In conclusion, a method combining parallel imaging, temporal low-rank constraints, and deep subspace learning spatial restoration was proposed to improve the image quality of highly accelerated cardiac parametric mapping sequences.

\section*{Acknowledgments}

This work made use of the Digital Research Alliance of Canada compute facilities. CS, JP, and GW receive funding from Canadian Institutes of Health Research grant number PJT178299. BK and MC receive funding from the Canada Research Chairs Program and NSERC Discovery grant number RGPIN/2023-03410.

\end{document}